%% file: main.tex
\documentclass[%
 aip,
prl,
twocolumn,
 amsmath,amssymb,
svgnames,
reprint,%
]{revtex4-2}

\usepackage{graphicx}
\usepackage{dcolumn}
\usepackage{bm}

\usepackage[utf8]{inputenc}
\usepackage[T1]{fontenc}
\usepackage{mathptmx}
\usepackage{etoolbox}
\usepackage{placeins}

\usepackage{lipsum}
\usepackage{placeins}
\usepackage{dashrule}
\usepackage{hyperref}
\usepackage{tikz}
\usepackage{blindtext}
\usepackage{times}
\usepackage{paralist}
\usepackage{ulem}
\usepackage{caption}
\usepackage{subcaption}

\newcommand{\crline}[1]{ \color{#1} $\bigcdot \,\bigcdot \,\bigcdot \,\bigcdot \,\bigcdot$}
\newcommand{\prline}[1]{\color{#1}\hdashrule[0.75ex]{0.97cm}{1pt}{1.3mm} }

\makeatletter
\newcommand*\bigcdot{\mathpalette\bigcdot@{.5}}
\newcommand*\bigcdot@[2]{\mathbin{\vcenter{\hbox{\scalebox{#2}{$\m@th#1\bullet$}}}}}
\makeatother

\makeatletter
\def\@email#1#2{%
 \endgroup
 \patchcmd{\titleblock@produce}
  {\frontmatter@RRAPformat}
  {\frontmatter@RRAPformat{\produce@RRAP{*#1\href{mailto:#2}{#2}}}\frontmatter@RRAPformat}
  {}{}
}%
\makeatother
\begin{document}

\preprint{AIP/123-QED}

\title[Indicator functions of normal turbulent stresses: log or quarter-power ]{Utilizing indicator functions with computational data to confirm nature of \\overlap in normal turbulent stresses: logarithmic or quarter-power}


\author{Hassan Nagib}%
\homepage{Author to whom correspondence should be addressed: nagib@iit.edu.}
\affiliation{%
ILLINOIS TECH (I.I.T), Chicago, IL, 60614 USA%
}%

\author{Ricardo Vinuesa}%
\homepage{rvinuesa@mech.kth.se}
\affiliation{%
FLOW, Engineering Mechanics, KTH Royal Institute of Technology, SE-100 44, Stockholm, Sweden
}%

\author{Sergio Hoyas}%
\homepage{serhocal@mot.upv.es}
\affiliation{%
 Instituto Universitario de Matem\'atica Pura y Aplicada, Universitat Polit\`ecnica de Val\`encia, Valencia 46022, Spain 
}%

\date{\today}

\begin{abstract}
Indicator functions of the streamwise normal-stress profiles (NSP), based on careful differentiation of some of the best direct numerical simulations (DNS) data from channel and pipe flows, over the range $550<Re_\tau<16,000$, are examined to establish the existence and range in wall distances of either a logarithmic-trend segment or a $1/4$-power region. For the nine out of fifteen cases of DNS data we examined where $Re_\tau<2,000$, the NSP did not contain either of the proposed trends. As $Re_\tau$ exceeds around $2,000$ a $1/4$-power, reflecting the ``bounded-dissipation'' predictions of Chen \& Sreenivasan \cite{che22,che23} and data analysis of Monkewitz \cite{monp23}, develops near $y^+=1,000$ and expands with Reynolds numbers extending to $1,000<y^+<10,000$ for $Re_\tau$ around $15,000$. This range of $1/4$-power NSP corresponds to a range of outer-scaled  $Y$ between around $0.3$ and $0.7$. The computational database examined did not include the zero-pressure-gradient boundary layer experiments at higher Reynolds numbers where the logarithmic trend in the NSP has been previously reported around $y^+$ of $1,000$ by Marusic et al. \cite{mar19,hwa22} according to a ``wall-scaled eddy model''.

\end{abstract}

\maketitle

\section{Introduction}
One of the most fascinating ideas about the behavior of turbulent flows is that even 140 years after the unofficial starting of this discipline with the article by Osborne Reynolds \cite{rey83}, many issues still attract a great deal of controversy. As turbulent flows are responsible for up to 15\% of the energy wasted by mankind \cite{jim13}, solving these problems is a matter of special urgency for the climate emergency. We believe no tools should be discarded \cite{lar23,vin22d,vin23d}. 

Two topics with considerable discussion in recent literature are related to the overlap regions in wall-bounded flows for the mean velocity profile (MVP) and the normal stresses (NSP). In the case of the MVP, the questions include: if pure-log overlap region or if the logarithmic plus linear (log+lin overlap) of Monkewitz \& Nagib \cite{mon23} is the more prevalent representation of the overlap, and if the non-universality of the overlap parameters including the K\'arm\'an ``constant'', have been sufficiently established or confirmed; see Nagib \& Chauhan \cite{variations} and Baxerras, Vinuesa \&Nagib \cite{bax24}. 

In the case of the NSP, the key questions we examine: is the trend of the turbulence normal stress, and in particular for the streamwise velocity, reflects for some wall distance and Reynolds number range the logarithmic behavior \cite{mar19,hwa22}~?, or does a $1/4$-power trend provide a better representation \cite{che22,che23,monp23}~?. 

The logarithmic trend results from models of wall turbulence based on a hypothesis of  ``wall-scaled eddies'' that grow linearly with inner-scaled wall distance and requires a universal K\'arm\'an constant, $\kappa$, leads to an infinite amplitude of the near-wall peak of the streamwise normal stress for the high $Re_\tau$ asymptotic limit, and predicts the existence of a second peak in the NSP \cite{sam18,hoy22}. An infinite growth of the peak in the NSP has been challenged in several ways and a brief description of the issues and our abilities to ever confirm the correct asymptotic limit unequivocally are described in a recent manuscript by Nagib, Monkewitz and Sreenivasan \cite{nag23}.

Chen \& Sreenivasan \cite{che23}, on the other hand, considered the inner law to be based on the peak value of NSP after normalization with the friction velocity, $u_\tau$. This is due to the fact that according to their theory, and in consideration of bounded dissipation, the peak values eventually saturate in terms of $u_\tau$, which is equivalent to simply normalizing by $u_\tau$. However, using the peak value takes into account finite Reynolds number effects, and the agreement with the data will be uniformly true for all Reynolds numbers. This is, in effect, their inner law.

They pick a standard and general relation for streamwise fluctuations in the outer flow to arrive at an overlap region in NSP. By matching with their inner law, described in more detail in their papers \cite{che22,che23} and by Monkewitz \cite{monp23}, one can show that any fluctuation depends on wall distance in the following universal manner:
\begin{equation}\label{eq:001}
    \phi(Y) = \alpha_\phi - \beta_\phi (Y)^{-1/4},
\end{equation}
which should apply in a region of the flow beyond the inner region, including the overlap region. Here, $Y=y/\delta$, where $\delta$ is the flow thickness (channel half-height, pipe radius, or boundary layer thickness), $\alpha_\phi$ and $\beta_\phi$ are constants, and $\phi$ represents any of the turbulence velocity components. For very large friction Reynolds number, $Re_\tau$, $\phi$ asymptotes to a constant, so this theory does not allow for a second peak.

Before continuing, we need to specify some notation. We will present some results for both channels and pipes. The streamwise, wall-normal, and spanwise (azimuthal) coordinates are $x$, $y$, and  $z (\theta)$. The corresponding instantaneous velocity components are $U_x$, $U_y$, and $U_z(U_\theta)$, respectively. Statistically averaged quantities in $x$ and $z (\theta)$ are denoted by angles, $\langle \phi \rangle$, whereas fluctuating quantities are denoted by lowercase letters, {\it i.e.}, $U_x=\langle U_x \rangle + u_x$. Primes are reserved for root mean squares (RMS) or intensities: $u'=\langle uu \rangle^{1/2}$. The friction Reynolds number $Re_{\tau}=\delta u_{\tau}/\nu$, where $\nu$ is the fluid kinematic viscosity.

Hassan Nagib was honored and delighted to present recent results on the overlap region of MVP for wall-bounded turbulence \cite{mon23} as the opening talk of ``Topics in Classical and Quantum Engineering Science Symposium Celebrating the career of K. R. Sreenivasan at 75!''\cite{Sre23}. A key to careful examination of such overlap regions is the use of indicator functions, which is based in this case on the derivative of the streamwise MVP. Summarizing here,  we expect that for a ``pure-log'' overlap region:
\begin{equation}\label{eq:002}
    \langle{U}^+_{x\rm OL}(y^+\gg 1~\&~Y\ll 1) \rangle = \frac{1}{\kappa} \ln y^+ + B.
\end{equation}
\
\par 
Here $y^+=yu_\tau/\nu$, where $\nu$ is the kinematic viscosity, and B is the interception constant.  If the MVP exhibits a ``log + lin" overlap region, the following trend would be representative of a range in wall distances between the inner part of the flow and the outer part that includes the ``wake'':
\begin{equation}\label{eq:003}
\langle{U}^+_{x\rm OL}(y^+\gg 1~\&~Y\ll 1) \rangle= \kappa^{-1}\ln y^+ + S_0  y^+ /Re_\tau + B_0 + B_1/Re_\tau.
\end{equation}

We also define $Y \equiv y^+ /Re_\tau$ to obtain the following from Equation~(\ref{eq:003}):
\begin{equation}\label{eq:004}
\langle{U_x}^+_{\rm OL}(Y\ll 1) \rangle\sim \kappa^{-1}\ln Y + \kappa^{-1}\ln Re_\tau +S_0  Y + B_0 + B_1/Re_\tau. 
\end{equation}
Simplifying, we obtain:
\begin{equation}\label{eq:005}
\langle{U_x}^+_{\rm OL}(Y\ll 1) \rangle= \kappa^{-1}\ln Y + \kappa^{-1}\ln Re_\tau +S_0  Y + B_0 + {\rm  H.O.T.,}
\end{equation}
where H.O.T. denotes higher order terms. In these equations, $S_0$, $B_0$, and $B_1$ are constants that need to be established either from experiments or DNS. To examine the MVP in the overlap region, the commonly used indicator function based on the mean velocity profile for wall-bounded turbulence, $\Xi$, can be obtained from:
\begin{equation}\label{eq:006}
 \Xi = y^+\frac{{\rm d}\left\langle {{U}_x^+}\right\rangle}{{\rm d}y^+} = Y\frac{{\rm d}\left\langle {{U}_x^+}\right\rangle}{{\rm d}Y}.   
\end{equation}
From Equations~(\ref{eq:005}) and (\ref{eq:006}), one can obtain the equation for $\kappa$ and $S_0$:
\begin{equation}\label{eq:007}
\Xi_{\rm OL} = \kappa^{-1} +S_0  y^+ /Re_\tau = \kappa^{-1} +S_0  Y .
\end{equation}

Fitting this linear equation to a selected range of $\Xi$ versus $Y$, the values of $\kappa$ and $S_0$ are easily extracted with the help of the indicator function from the slope and intercept, respectively.  Based on various tests including by Monkewitz \& Nagib \cite{mon23}, we utilize the range in $Y$ between $0.3$ and $0.6$ for all Reynolds numbers.

In the current work, we focus on the streamwise normal stress and utilize a similar indicator function, $\zeta_{uu}$, defined as:
\begin{equation}\label{eq:008}
 \zeta_{uu} = y^+\frac{{\rm d}{\left\langle u_x^+u_x^+\right\rangle}}{{\rm d}y^+} = Y\frac{{\rm d}{\left\langle u_x^+u_x^+\right\rangle}}{{\rm d}Y}.   
\end{equation}

Finally, to examine if the $1/4$-power trend is reflected in the normal stress profiles, a complementary indicator function, $\zeta_{uu,BD}$, based on the bounded-dissipation predictions of Equation~(\ref{eq:001}), is defined by:
\begin{equation}\label{eq:009}
 \zeta_{uu,BD} = 4Re_\tau^{1/4}(y^+)^{3/4}\frac{{\rm d}{\left\langle u_x^+u_x^+\right\rangle}}{{\rm d}y^+} = 4Y^{3/4}\frac{{\rm d}{\left\langle u_x^+u_x^+\right\rangle}}{{\rm d}Y}.   
\end{equation}

In the case of the of the indicator functions of the NSP, the range we examine the applicability of either Equation~(\ref{eq:008}) or (\ref{eq:009}) is not set and its lower and upper limits may depend on $Re_\tau$. In a plot of $\zeta_{uu}$ versus wall distance, a constant horizontal segment or plateau that expands with $Re_\tau$ represents a logarithmic trend within the normal stress profiles. On the other hand, a constant horizontal segment or plateau in plots of $\zeta_{uu,BD}$ that expands with $Re_\tau$ would represent a $1/4$-power trend for the corresponding part of the NSP.

For indicator functions such as $\Xi$, $\zeta_{uu}$ and $\zeta_{uu,BD}$, the required differentiation of the profiles with sparse, unequally distributed, or limited-accuracy data, is a big limitation that leads to results with low confidence levels and higher ambiguity. Therefore, unlike in the efforts of Monkewitz \& Nagib \cite{mon23} and Baxerras et al.\cite{bax24}, we have relied here exclusively on direct numerical simulations (DNS) data. Among the three main classes of wall-bounded flows, only channel and pipe flows are included in this study. We have not included any boundary layer flows due to challenges in boundary conditions, specifically away from the wall (the ``top'' boundary condition) and the role played by intermittency in defining the limits of the overlap region. We have selected the best cases available to us that are listed in Table~\ref{tab:1} with the type and color representing them in all figures.

\input{table1}

The recent results by Monkewitz \& Nagib \cite{monp23} revealing the addition of a linear term in the overlap region of wall-bounded flows, its wider and better-defined range, and its location farther from the inner flow, have three important advantages we have recently demonstrated at two conferences in Texas A\&M \cite{Sre23}  and KAUST\cite{kau24}:\\
\begin{enumerate}
    \item Ability to arrive at the asymptotic values of the parameters of the overlap region at lower Reynolds numbers,
    \item Utilizing the indicator function of the MVP, it is far easier to identify the overlap region and to extract $\kappa$ and $S_0$ using Equation~(\ref{eq:007}) than to identify a plateau for a pure log region, and
    \item With the typical uncertainty of available data, the best-fit overlap-region parameters are arrived at more easily with higher accuracy as compared to selecting between pairs of $\kappa$ and $B$ for the pure-log overlap of Equation~(\ref{eq:002}).
\end{enumerate}

\begin{figure}
   \centering
      \includegraphics[width=0.49\textwidth]{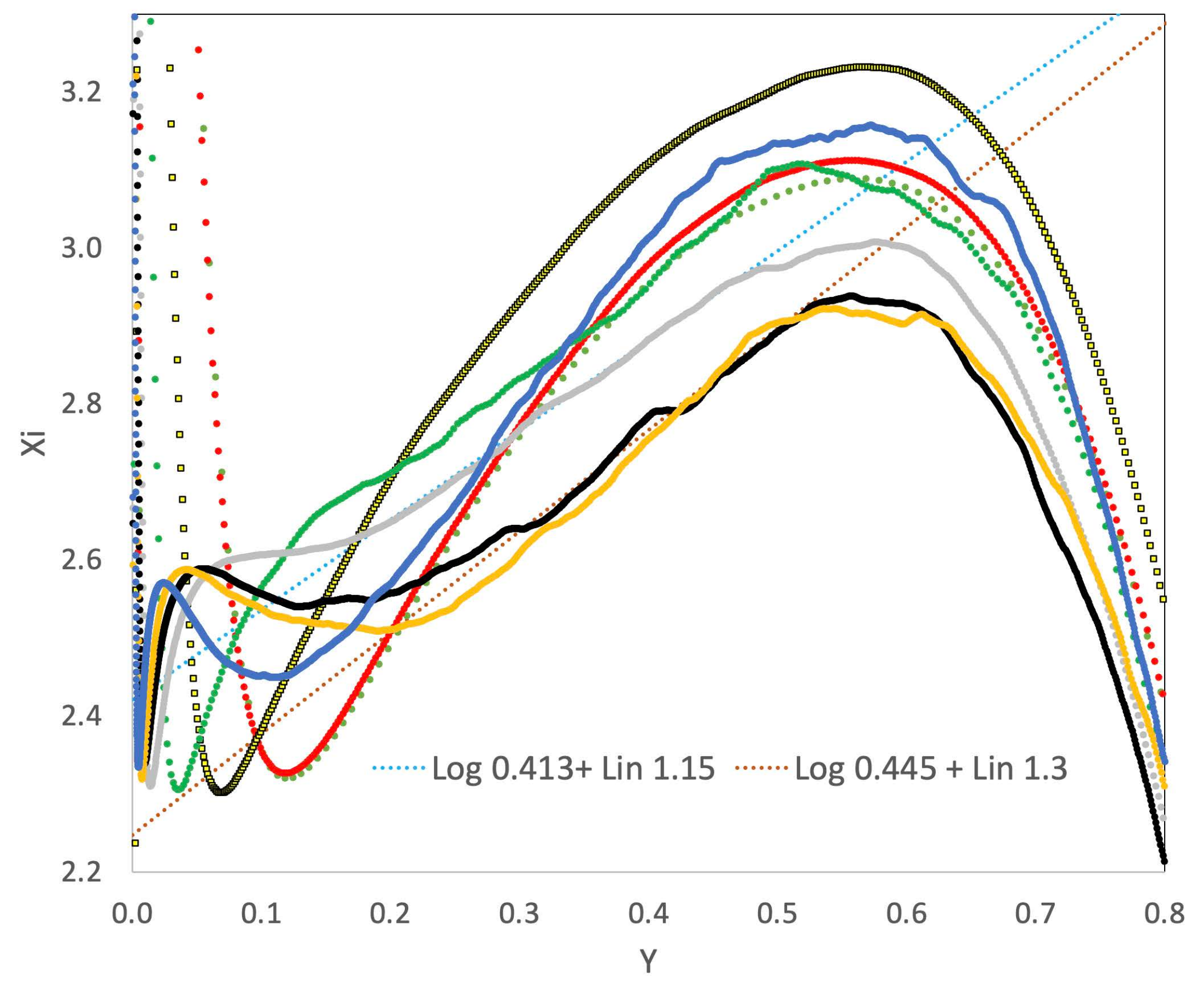}
    \caption{Indicator function of the MVP, $\Xi=y^+\frac{\rm d}{{\rm d} y^+}\langle{U}^+\rangle=Y\frac{\rm d}{{\rm d} Y}\langle{U}^+\rangle$, for channel flows with $550<Re_\tau<10,000$. Lines, symbols, and colors are described in Table~\ref{tab:1}, and the dotted lines represent two fits of overlap region of $\Xi$ by logarithmic plus linear relation of Equation~(\ref{eq:007})}
    \label{fig:fig1}
\end{figure}

Figure~\ref{fig:fig1} displays typical results for the MVP indicator function as a reference to compare with the same cases for the NSP indicator functions. Recently \cite{hoy23b}, we systematically examined some of the resolution and convergence requirements for accurate determination of the log+lin extended overlap region in wall-bounded turbulence. Some of these effects are reflected in Figure~\ref{fig:fig1}, including the impact of the different resolution and convergence on the extracted values of $\kappa$ and $S_0$. The better-resolved and converged case at the intermediate $Re_\tau$ of $5,200$ by Lee \& Moser \cite{lee15} represents the best agreement with experiments \cite{mon23}.

\begin{figure}
    \centering
        \includegraphics[width=0.49\textwidth]{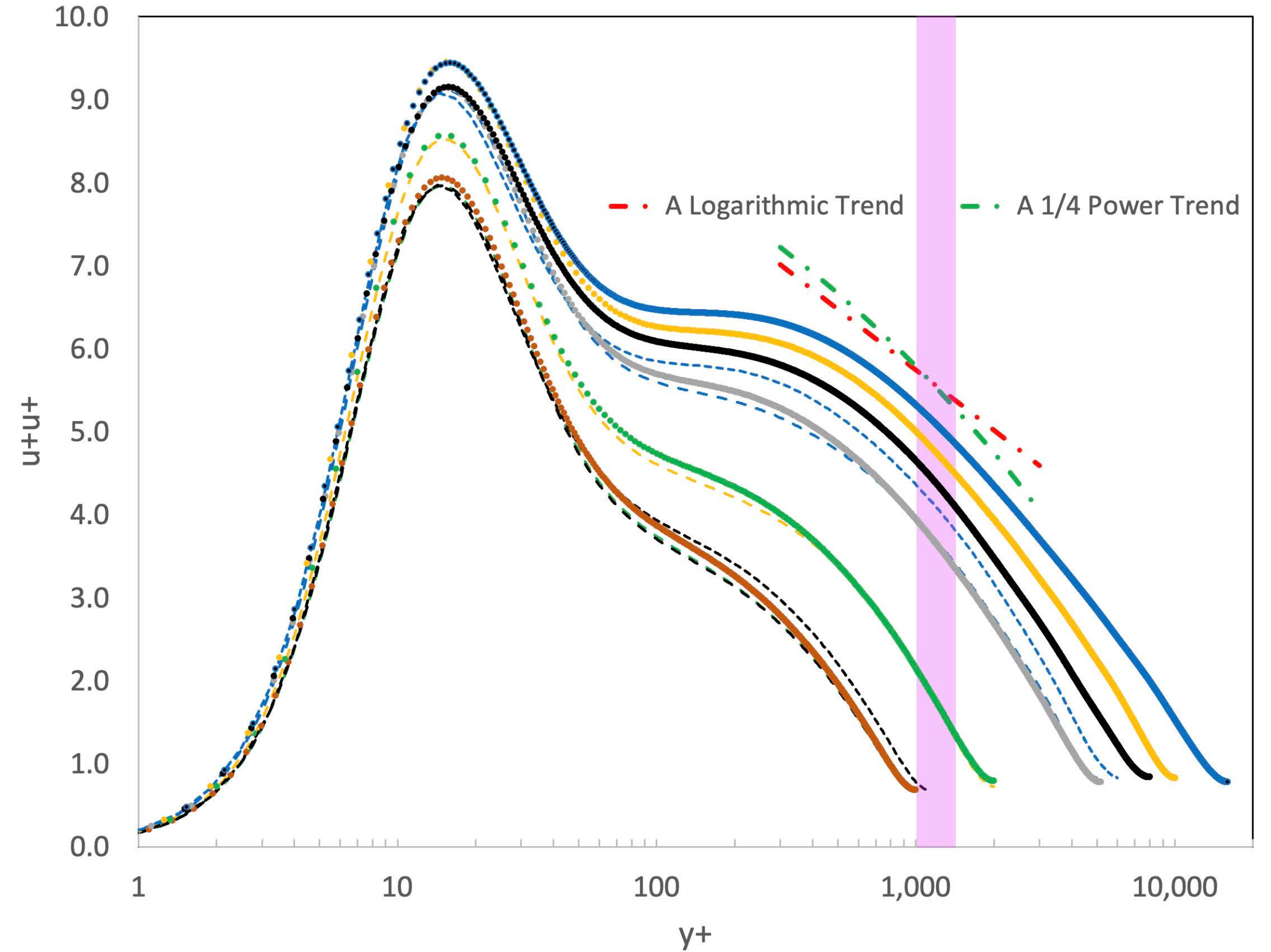}
     \caption{Streamwise normal stress distribution in inner-scaled variables for channel and pipe flows with $1,000<Re_\tau<16,000$. The two dash-dot lines represent the contrast between the logarithmic and $1/4$-power trends as a potential representation of an overlap region near the pink shaded range cantered at $y^+=1,200$.  Lines, symbols, and colors are described in Table~\ref{tab:1}.}     
    \label{fig:fig2}
\end{figure}

\section{Results and Discussion}

In Figure~\ref{fig:fig2}, a collection of NSP from most of the cases of Table~\ref{tab:1} is presented.  To contrast differences between the two proposed trends in some segments of the profiles, representative logarithmic and $1/4$-power trends are depicted on the figure around a potential applicable region around $y^+$ of nominally $1,200$.  The parameters of both are arbitrarily selected for this figure to demonstrate the contrast between them. Such parameters, including the location and range of best fit, would depend on $Re_\tau$. The sample trends shown here and the choice of the location of the vertical purple line are selected with a focus on cases with $Re_\tau > 2,000$; i.e., the curves of NSP to the right of the green one.  The figure demonstrates the ability of the $1/4$-power trend to accommodate the changing slope of the NSP even for $Re_\tau$ as low as $2,000$ for both channel and pipe flows.

\begin{figure}
      \centering
      \begin{subfigure}{0.49\textwidth}
      \includegraphics[width=0.95\textwidth]{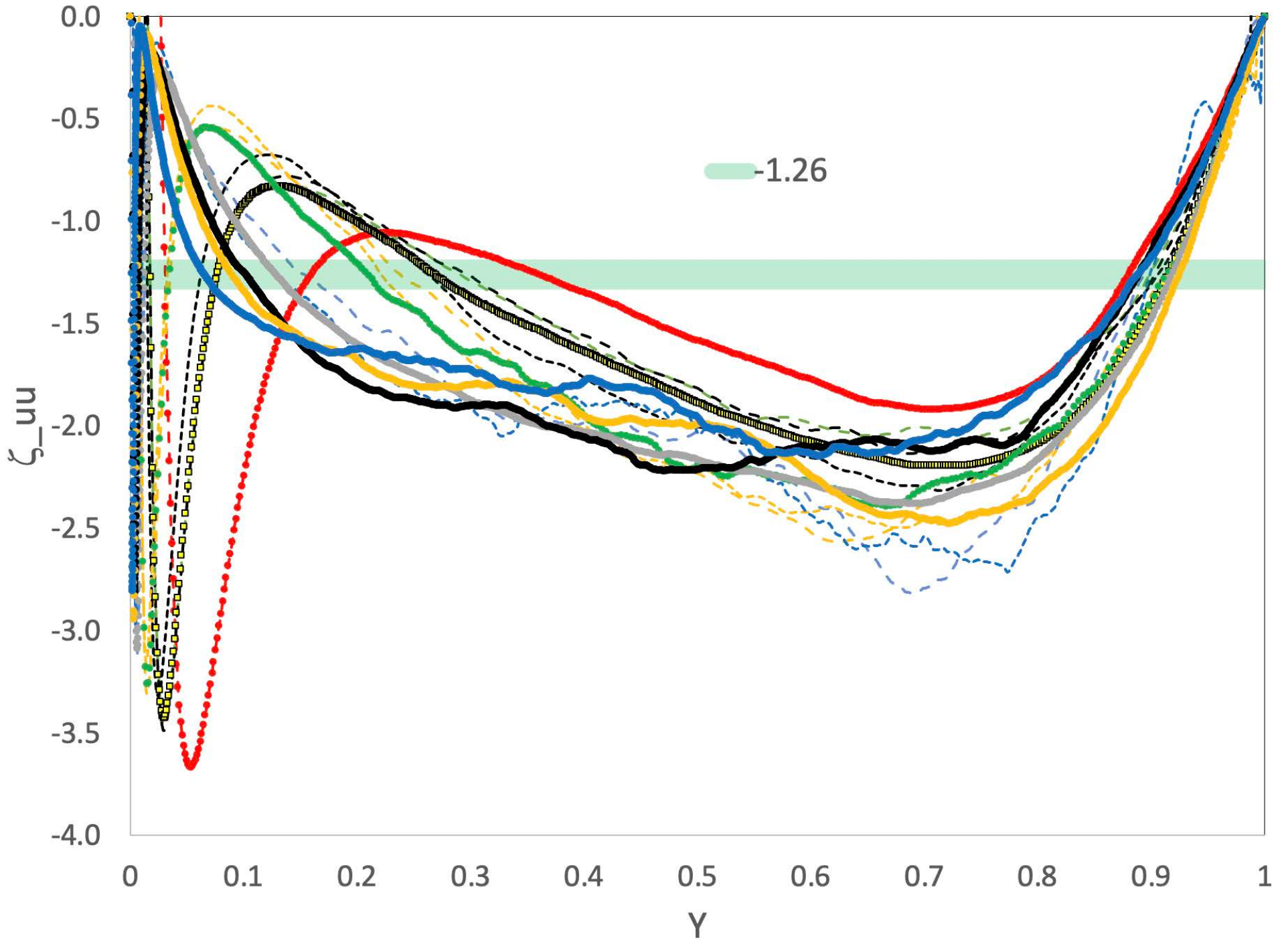}
     \end{subfigure}
      \begin{subfigure}{0.49\textwidth}
      \includegraphics[width=0.95\textwidth]{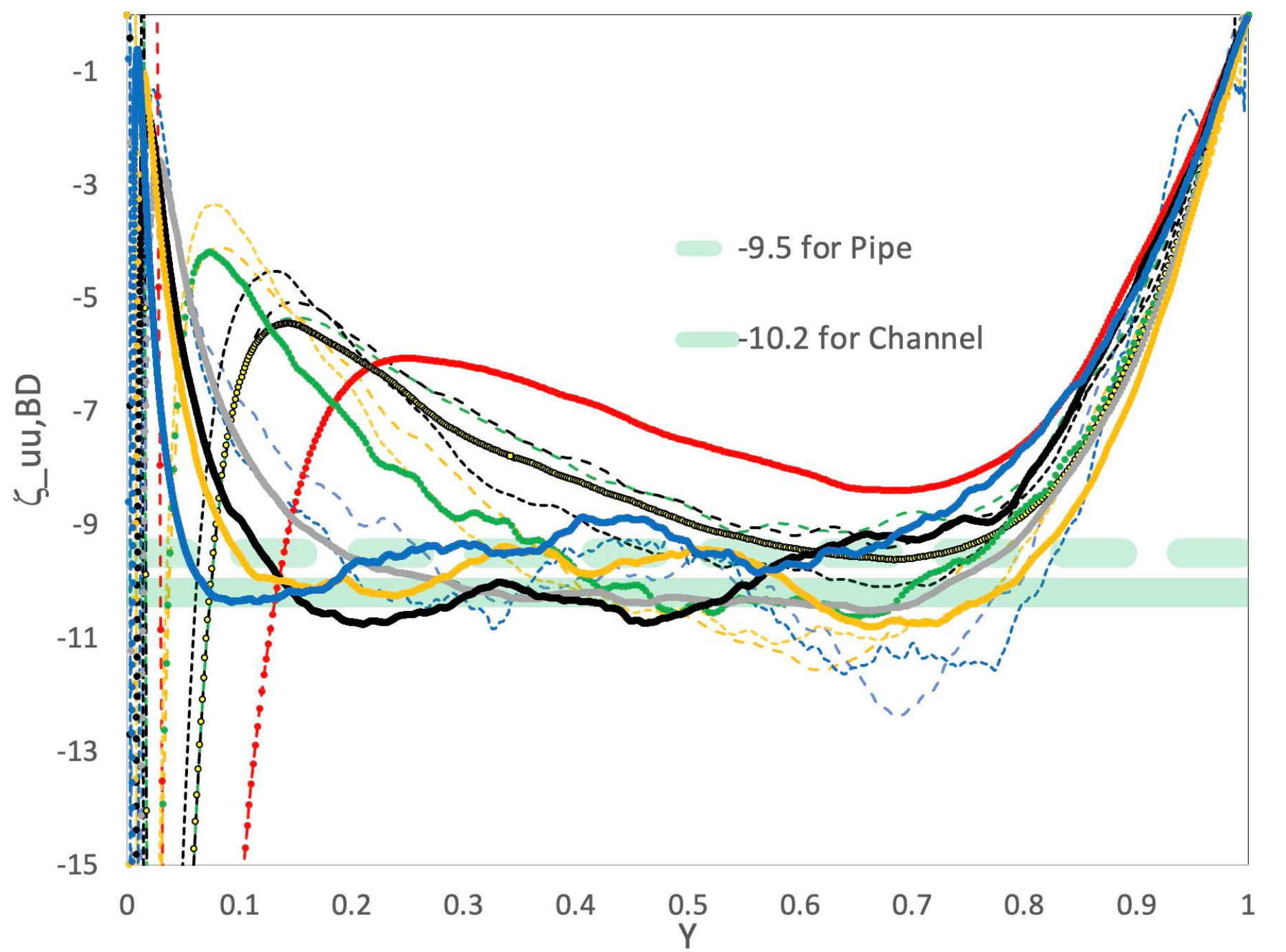}
     \end{subfigure}
     \caption{\label{fig:fig3} Indicator functions of streamwise normal stress versus outer-scaled wall distance $Y$ for channel and pipe flows with $550<Re_\tau<16,000$.  Top: Indicator function $\zeta_{uu}=Y\frac{\rm d}{{\rm d} Y}\langle{u^+u^+}\rangle$ for logarithmic trend identification. Bottom: Bounded-dissipation corrected indicator function $\zeta_{uu,BD}=4Y^{3/4}\frac{\rm d}{{\rm d} Y}\langle{u^+u^+}\rangle$ for $1/4$-power trend identification. Lines, symbols, and colors are described in Table~\ref{tab:1}, and horizontal green lines represent some predictions.}
\end{figure}

\begin{figure}
      \centering
      \begin{subfigure}{0.49\textwidth}
      \includegraphics[width=0.95\textwidth]{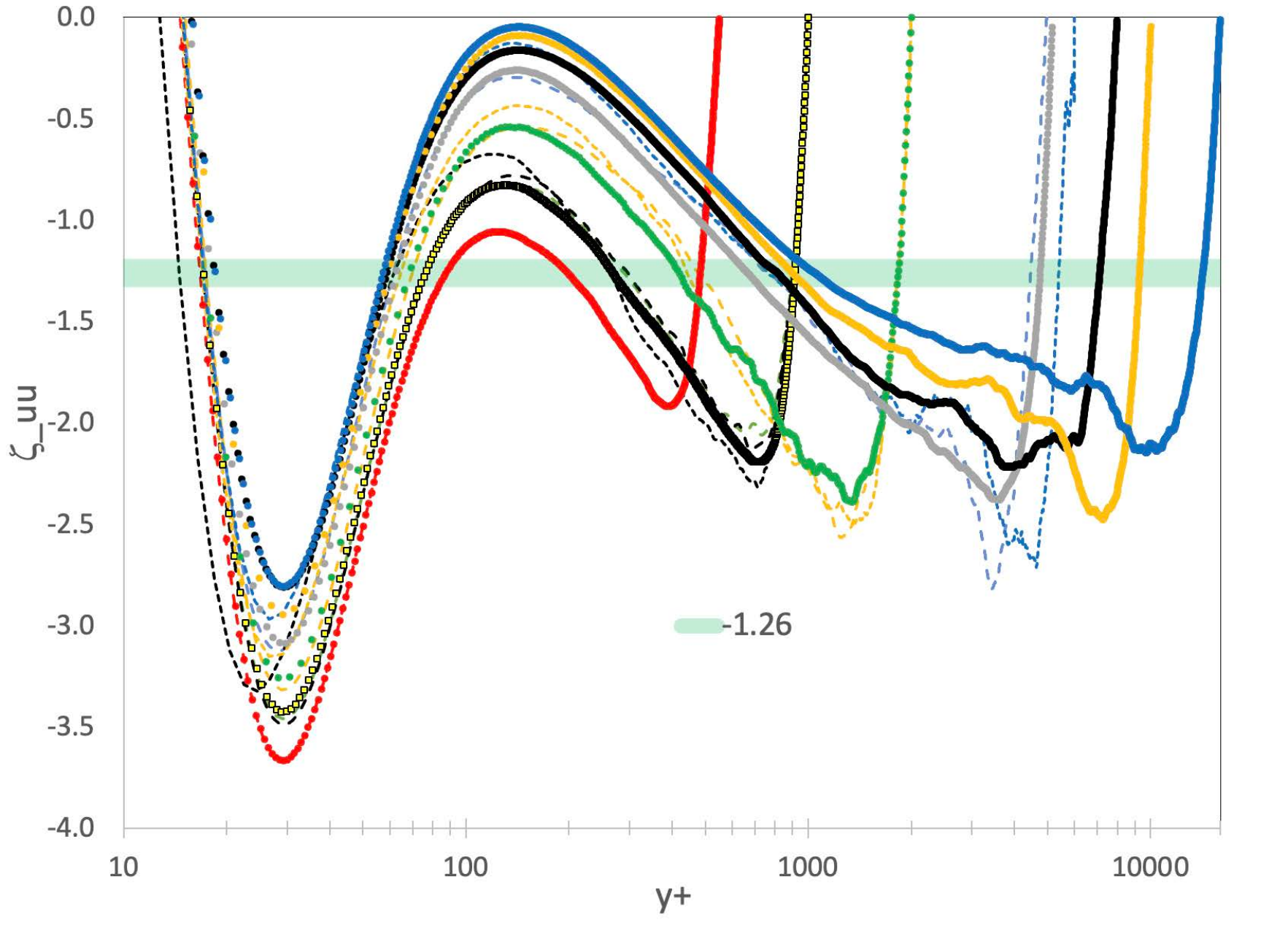}
     \end{subfigure}
      \begin{subfigure}{0.49\textwidth}
      \includegraphics[width=0.95\textwidth]{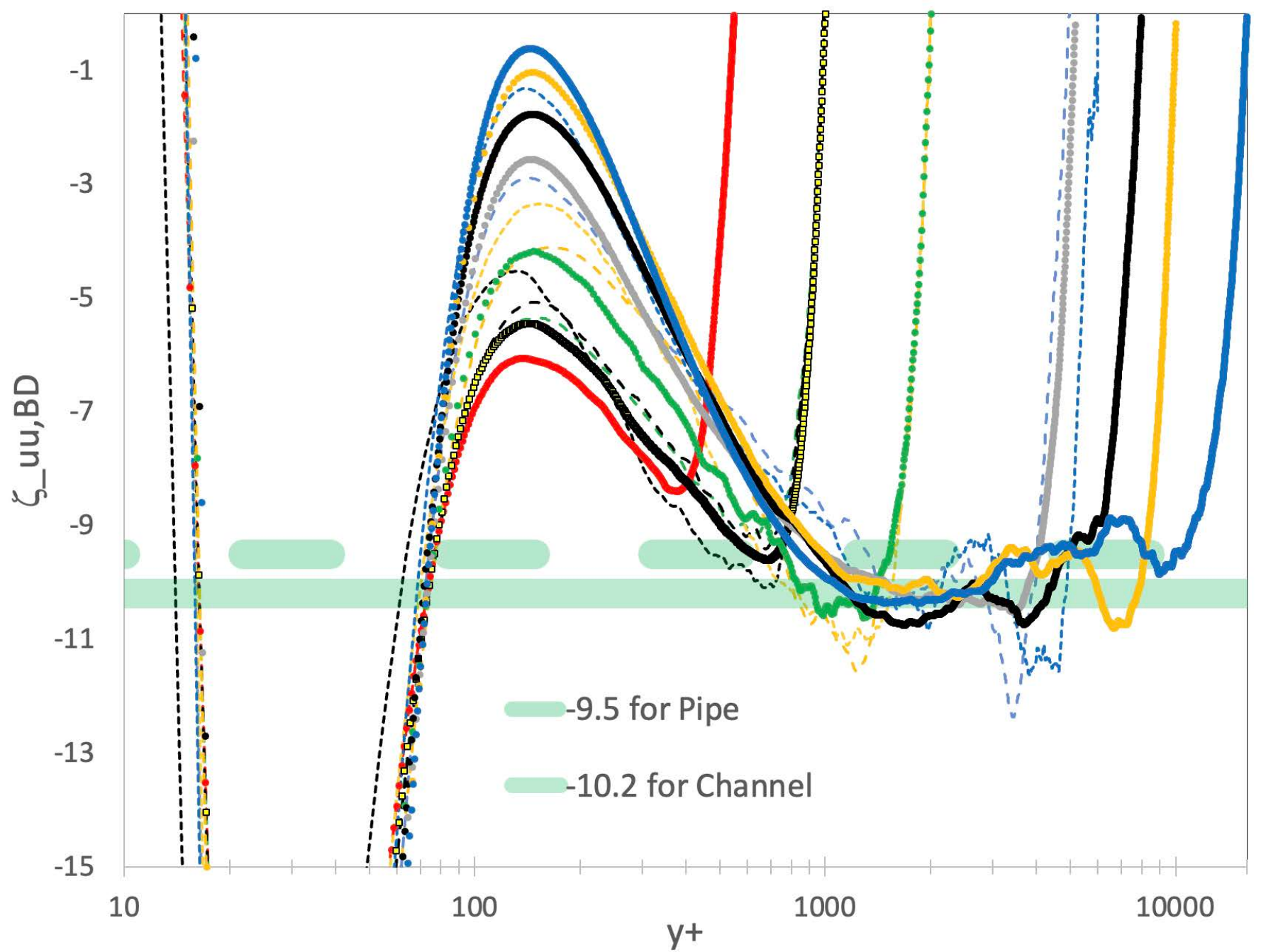}
     \end{subfigure}
    \caption{\label{fig:fig4} Same as Figure~\ref{fig:fig3}, but as a function of the inner-scaled wall distance $y^+$.}
     
\end{figure}

Figures~\ref{fig:fig3} and \ref{fig:fig4} include the results from all the cases of Table~\ref{tab:1} for both channels and pipes. The top part of both figures presents the indicator function, $\zeta_{uu}$, focused on revealing a logarithmic region. The bottom parts are intended to identify, using the complementary indicator function of Equation~(\ref{eq:009}), $\zeta_{uu,BD}$, the existence and range of the $1/4$-power trend.  Figure~\ref{fig:fig3} presents the respective indicator functions as a function of outer-scaled wall distance $Y$, and Figure~\ref{fig:fig4} presents the data versus the inner-scaled wall distance $y^+$.  The horizontal green wide lines indicate expected levels of the constant parameters of the two fits. For the top parts of the figures, the often-used value of $-1.26$ \cite{mar19,hwa22} is represented. For the indicator function focused on the bounded dissipation with a $1/4$-power prediction, the values recently reported by Monkewitz \cite{monp23} are displayed for channel and pipe flow as $-10.2$ and $-9.5$, respectively.

These two figures reveal a $1/4$-power range, as reflected by horizontal near-flat segments in the bottom parts of the figures, and an absence of logarithmic trends that would have been represented by plateaus in the top parts of the figures is observed.  From the bottom part of Figure~\ref{fig:fig4}, the horizontal segments start appearing at around $y^+=1,000$ for $Re_\tau=1,000$ and extend to larger $y^+$ locations with $Re_\tau$, for both channel and pipe flows. Within the accuracy of the results in the bottom part of Figure~\ref{fig:fig3}, the range of the indicator function $\zeta_{uu,BD}$ displaying the $1/4$-power behavior is $0.3<Y<0.7$ for $Re_\tau>2,000$.

Examining the trends of Figure~\ref{fig:fig3} and Figure~\ref{fig:fig4}, it appears that pipe flows may require higher $Re_\tau$ conditions before the $1/4$-power trend is established. With better resolved and converged pipe flow DNS using higher order methods in adequate domain sizes, this observation may turn out to be premature.

\begin{figure}
      \centering
      \begin{subfigure}{0.49\textwidth}
      \includegraphics[width=0.9\textwidth]{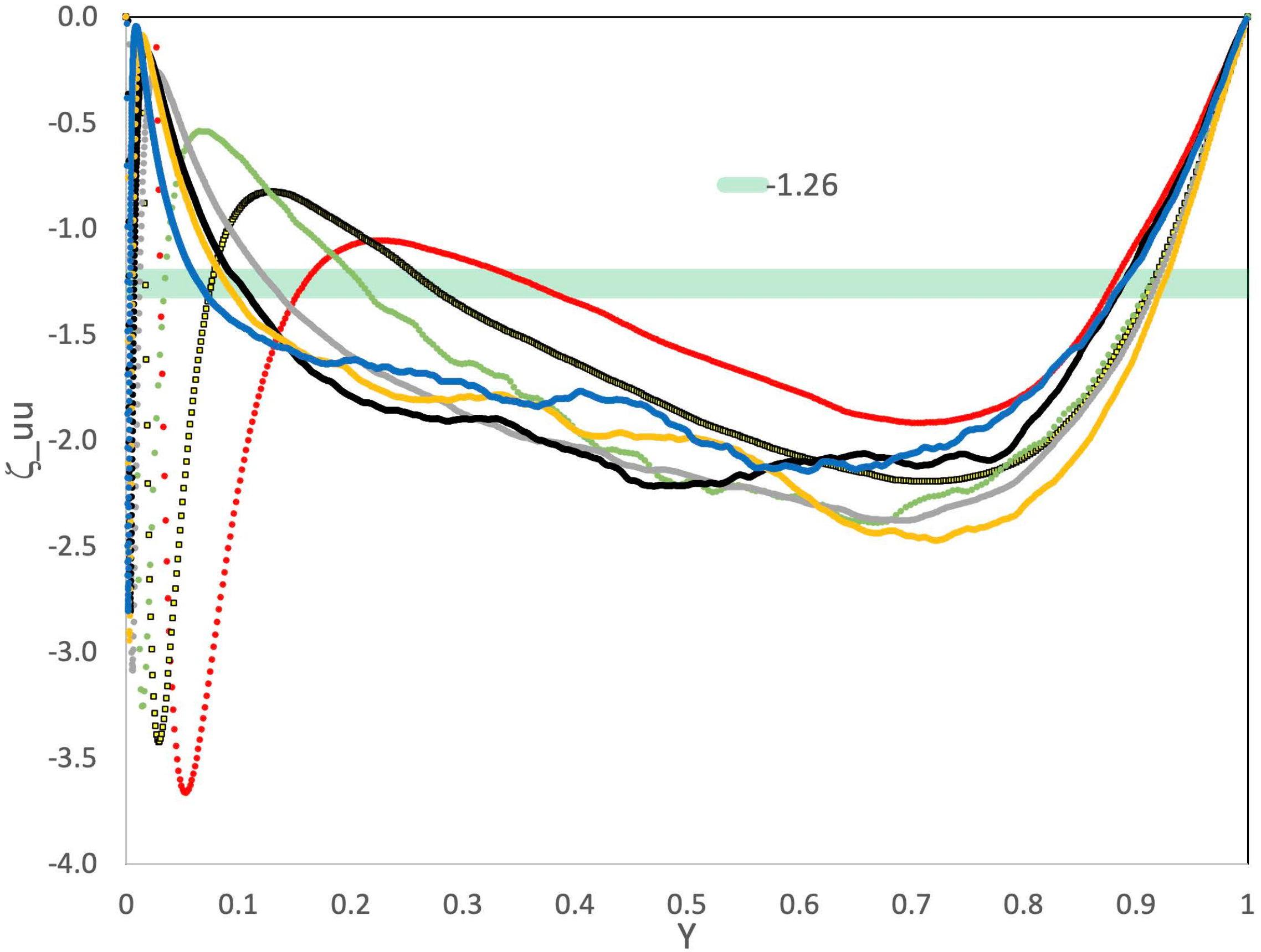}
     \end{subfigure}
      \begin{subfigure}{0.49\textwidth}
      \includegraphics[width=0.9\textwidth]{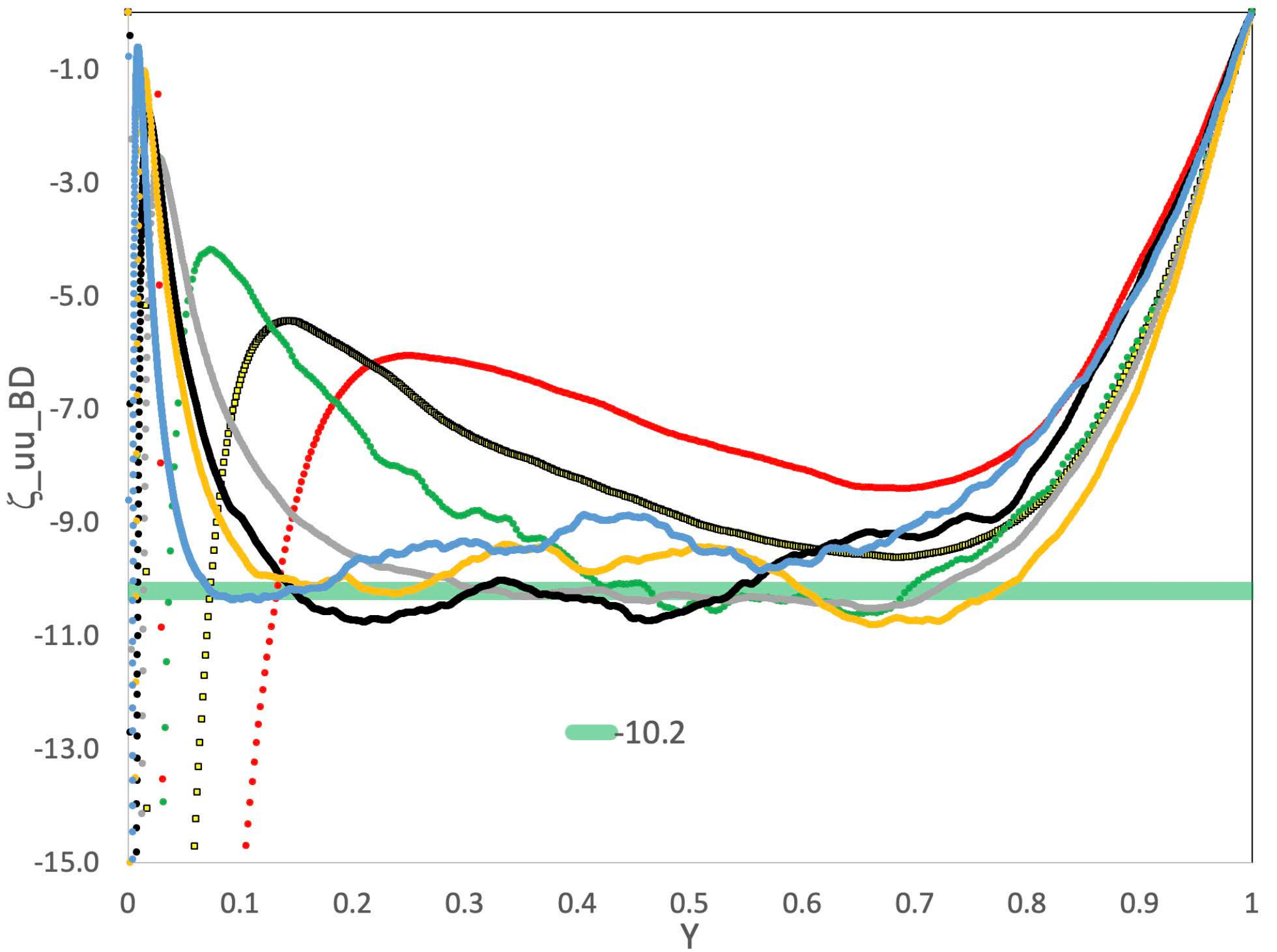}
     \end{subfigure}
     \caption{\label{fig:fig5} Same as Figure~\ref{fig:fig3}, but here we show only the channel flow data}
\end{figure}

We then focus on the channel flow cases in Figure~\ref{fig:fig5} because of the wider range of available Reynolds numbers.  The observations made based on Figure~\ref{fig:fig3} and Figure~\ref{fig:fig4} more clearly demonstrated and confirmed.  It is interesting to observe in the bottom part of Figure~\ref{fig:fig5} that the first appearance and development of the $1/4$-power region starts with the case of $Re_\tau=2,000$ (green line) the higher values of $0.45<Y<0.7$ extending down to $Y=0.3$ for $Re_\tau=10,000$, and possibly even a lower value for the two cases by Yamamoto et al. \cite{kan21,yam24}.  Concerns about the second order of the finite difference scheme used in these cases and possible limited convergence of the results may be contributing factors, limiting us from concluding the extension of the $1/4$-power range to outer-scaled wall distances below $Y<0.3$. Therefore, in the last part of the discussion, we study more carefully the potential extent of the $1/4$-power trend in both inner- and outer-scaled wall distances.

\begin{figure}
      \centering
      \begin{subfigure}{0.49\textwidth}
      \includegraphics[width=0.95\textwidth]{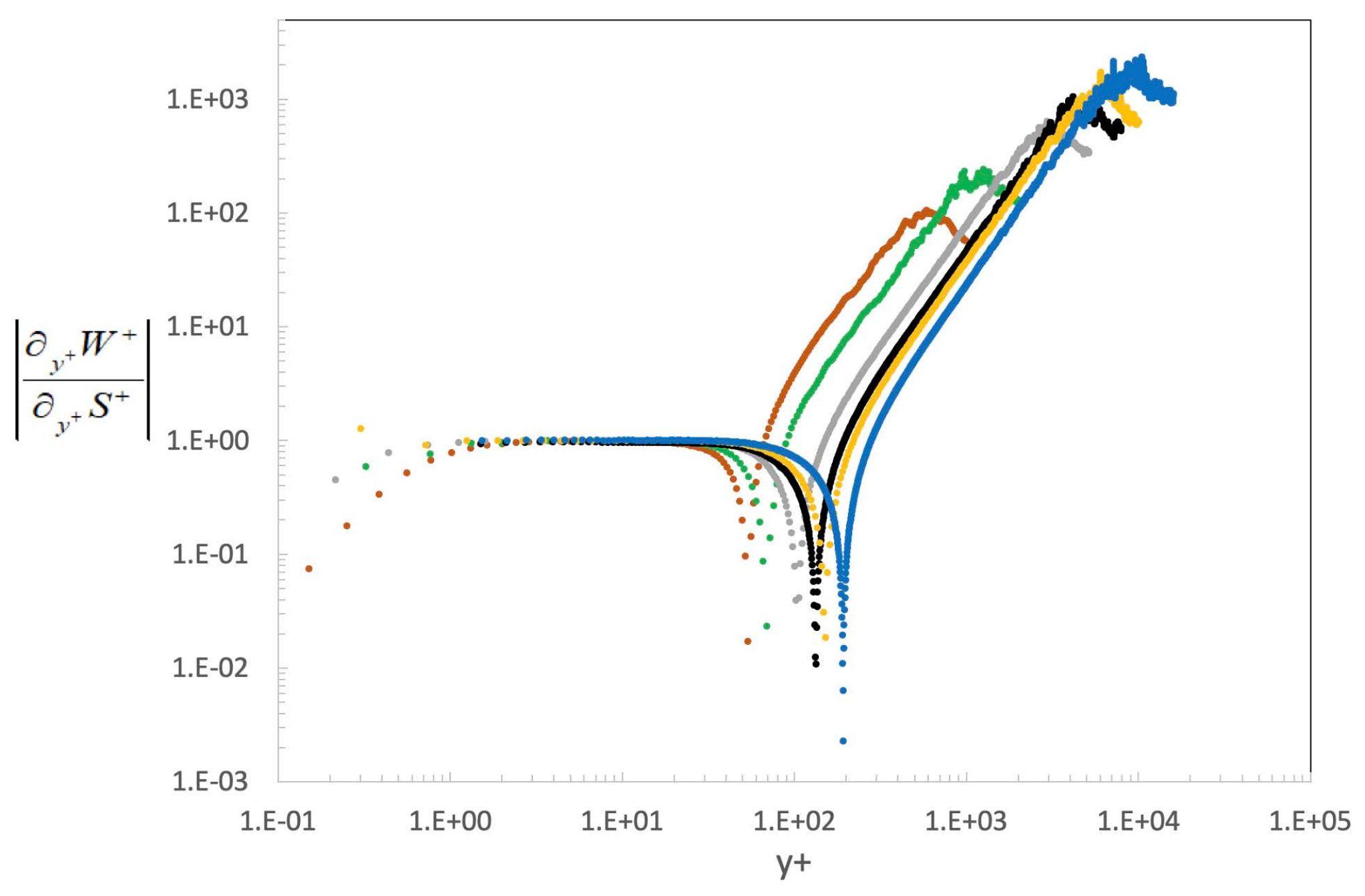}
     \end{subfigure}
      \begin{subfigure}{0.49\textwidth}
      \includegraphics[width=0.95\textwidth]{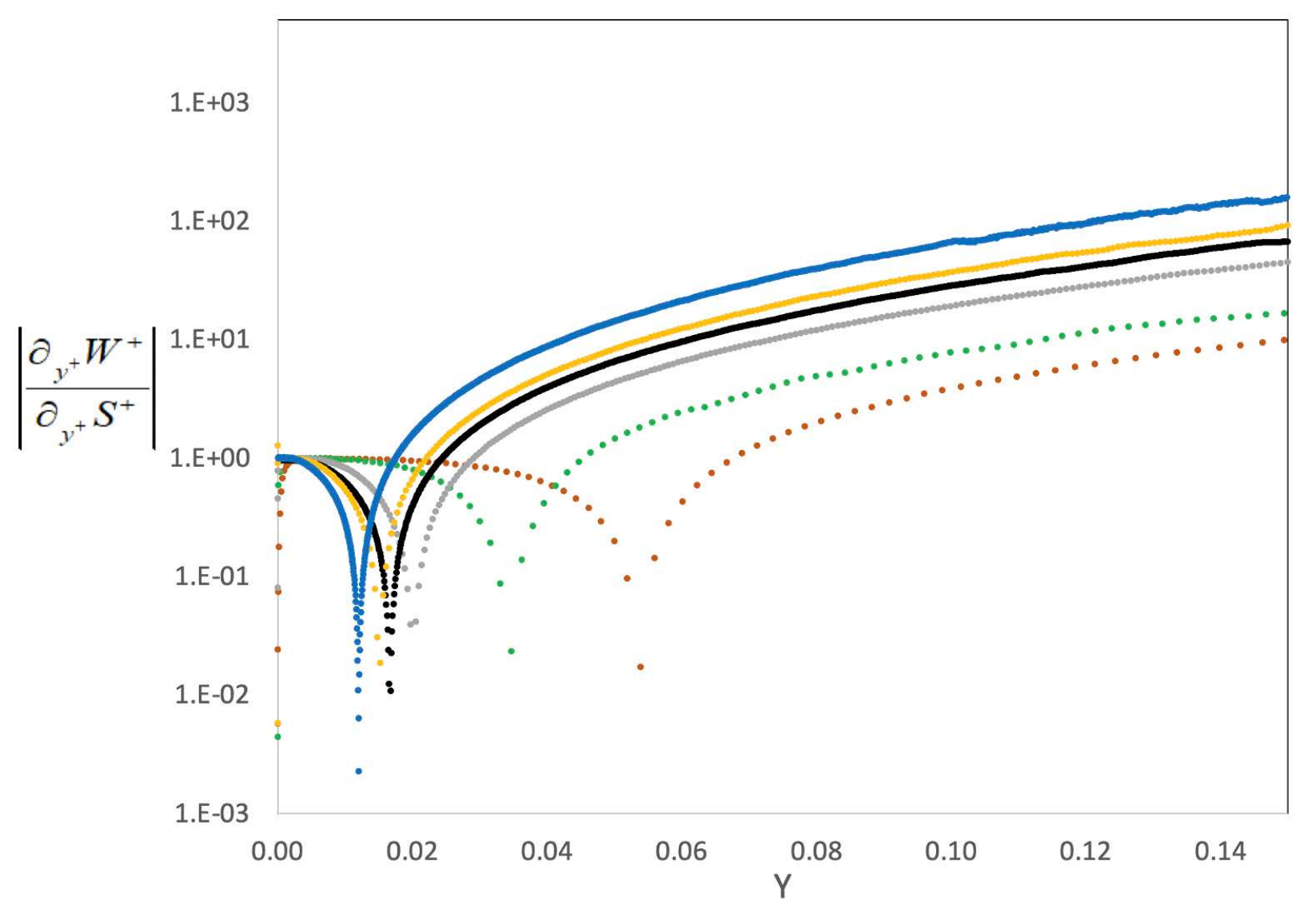}
     \end{subfigure}
     \caption{\label{fig:fig6} Turbulent stress gradient divided by viscous stress gradient displayed versus the wall distance in channel flows with $1,000<Re_\tau<16,000$. Top: presented with logarithmic scale against the inner-scaled wall distance $y^+$. Bottom: plotted with linear scale against the outer-scaled wall distance $Y$. Lines, symbols, and colors are described in Table~\ref{tab:1}}
     
\end{figure}

During K. R. Sreenivasan's presentation in the session on ``Evolution of the turbulent stresses with Reynolds number'' at the recent workshop held in KAUST\cite{kau24}, he displayed a version of the top part of Figure~\ref{fig:fig6}, with fewer Reynolds numbers. Inspired by his comments during the presentation, we expand on his figure here with higher $Re_\tau$ cases. In addition, we present the same data in the bottom part of the figure, but displayed versus the outer-scaled wall distance $Y$.  Both parts of Figure~\ref{fig:fig6} present the ratio of the turbulent stress gradient in the wall-normal direction divided by the viscous stress gradient. also in the wall-normal direction, as defined by:   
\begin{equation}\label{eq:010}
\left| \frac{\frac{{\rm d} W^+}{{\rm d} y^+}}{\frac{{\rm d} S^+}{{\rm d} y^+}} \right| = \left| \frac{{\rm d} \left\langle -u_x u_y \right\rangle}{{\rm d}y^+} / \frac{{\rm d^2}  U_x }{{\rm d}(y^+)^2}\right|   .
\end{equation}

Comparing the two panels of Figure~\ref{fig:fig6} we observe that the demarcation between the near wall region with an almost equal balance between the turbulent and viscous stress gradients and the overlap region, separated by the location of the minimum peak, shifts to higher values of $y^+$ with $Re_\tau$ in the top part of figure, while shifting to lower values in $Y$ with increasing Reynolds numbers as depicted in the bottom part of Figure~\ref{fig:fig6}. Further work is required to clarify whether the trends in the bottom part of Figure~\ref{fig:fig5} for some of the higher $Re_\tau$ cases in channel flows can be verified based on the trends of Figure~\ref{fig:fig6}.

\section{Conclusions}
The current work focused on utilizing the indicator function of the streamwise normal-stress profiles (NSP), $\zeta_{uu}$, of Equation~(\ref{eq:008}) using high-order differentiation of some of the best DNS data from channel and pipe flows.  The data includes fifteen different cases over the range $550<Re_\tau<16,000$. A constant value over some range of the wall-normal direction in $\zeta_{uu}$ would represent a logarithmic behavior consistent with predictions of the ``wall-scaled eddy hypothesis''\cite{mar19,hwa22}. A complementary indicator function of the streamwise normal-stress profiles (NSP), $\zeta_{uu,BD}$, based on Equation~(\ref{eq:009}) was also used to establish the existence and range in wall distances of a $1/4$-power region in the NSP. 

For $Re_\tau<2,000$, the NSP do not contain either of the proposed trends in the entire data range. As $Re_\tau$ exceeds around $1,000$, a $1/4$-power, reflecting the ``bounded dissipation'' predictions of Chen \& Sreenivasan \cite{che22,che23} and data analysis of Monkewitz \cite{monp23}, develops near $y^+=1,000$ and expands with Reynolds numbers extending to $1,000<y^+<10,000$ for $Re_\tau$ around $15,000$. This range of $1/4$-power in the NSP corresponds to a range of outer variable $Y$ between around $0.3$ and $0.7$. 

The database examined did not include the zero-pressure-gradient-boundary layers at higher Reynolds numbers where the logarithmic trend in the NSP has been previously reported around $y^+$ of $1,000$ by Marusic et al. \cite{mar19,hwa22} according to ``wall-scaled eddy model''. Recent results about the non-universality of the K\'arm\'an ``coefficient'', $\kappa$, in a wide range of pressure gradient boundary layers \cite{bax24}, is at odds with the requirement for a universal $\kappa$, for consistency with the logarithmic trend in the NSP. The true zero-pressure-gradient turbulent boundary layer, where Ivan Marusic and his group found the best evidence for a logarithmic trend in the NSP could be the exception, having a well established $\kappa=0.384$ \cite{mon23}.

Finally, the generally believed idea that the mean flow converges sooner than the fluctuations in turbulence computations appears to be contradicted by comparing results like those of Figure~\ref{fig:fig1} for the indicator functions of MVP at different Reynolds numbers and the indicator functions of the NSP in Figures~\ref{fig:fig3}, \ref{fig:fig4} \& \ref{fig:fig5}.  For example, the development of the $\zeta_{uu}$ with $Re_\tau$ in the top part of Figure~\ref{fig:fig4} is more orderly, and the agreement of $\zeta_{uu,BD}$ among a wide range of Reynolds numbers in the overlap region of the $1/4$-power trend in the bottom part of the same figure, are far more consistent than the observed trends of $\Xi$ in Figure~\ref{fig:fig1}.

\begin{acknowledgments}
Hassan is grateful for the near half a century of close friendship with Sreeni. They were introduced to each other by some of their mentors, including Stan Corrsin and Mark Morkovin and have continued to be near-brothers all these years although coming from different cultures.  

All authors would like to wish Sreeni many happy returns, and thank all the authors who share their data with them, both privately and publicly.  

SH was funded by project PID2021-128676OB-I00 by MCIN/AEI/10.13039/501100011033 and by “ERDF A way of making Europe”, by the European Union (SH). RV acknowledges the financial support from ERC grant no. `2021-CoG-101043998, DEEPCONTROL'. Views and opinions expressed are however those of the author(s) only and do not necessarily reflect those of the European Union or the European Research Council. Neither the European Union nor the granting authority can be held responsible for them. 

\end{acknowledgments}

\section*{Data Availability Statement}
The data that support the findings of this study are available from the authors upon reasonable request. The data referenced in Table I is available through the databases created by the authors of these papers. 
\bibliographystyle{unsrt}
\bibliography{turbulenceHoyas}

\end{document}

%% file: table1.tex
\begin{table*}[t]
	\begin{center}
	\begin{tabular}{lcccccccccccccc}
  		\underline{}Case & Type & Line  & $Re_{\tau}$ & $L_x/\delta$  & $L_z/\delta$ & $\Delta x^+$ & $\Delta z^+$ &$\min(\Delta y^+)$ & $\max(\Delta y^+)$& $ N_x $ & $ N_y $ & $ N_z $ & $T u_\tau/ h$ & Method \\ \hline
     
     	CHV005 \cite{hoy23b} & Channel & \crline{red} & 550 & $10\pi$ & $2\pi$ & $9$ & $4.5$ &0.05 & 1.68 &  1,536 & 901 & 1,152 & 150 & PS + CFD \cite{llu21c} \\
        PHV005 \cite{hoy23b} & Pipe & \prline{red} & 550 & $8\pi$ & $3\pi$ & $5.62$ & $3$ & $0.02$ & $3.2$  & 3,072 & 512 & 1,152 & 87 & PS + HOFD \cite{wil17}\\
         CHV010 \cite{hoy23b} & Channel & \crline{brown} & 1,000 & $8\pi $ & $3\pi $ & $8.1$ & $4.1 $ & 0.075 &2.5 & 3,072 & 1,085 & 2,304 & 30.7 &  PS + CFD \cite{llu21c}  \\
         PHV010 \cite{hoy23b} & Pipe & \prline{green} & 1,000  & $10\pi $ & $2\pi $ & $10.2$ & $2.72$ & 0.015 & 3.92 &  3,072 & 3,084 & 2,308 & 38 & PS + HOFD \cite{wil17}\\
         PYR010 \cite{yao23} & Pipe & \prline{black} & 1,000  & $10\pi$ & $2\pi $ & $10.2$ & $3.9$ & 0.1 & 4.9 & 3,072 & 384 & 1,280  & 17.1 & PS + HOFD \cite{wil17}\\
        PPR010 \cite{pir21} & Pipe & {\color{black}\hdashrule[0.75ex]{0.86cm}{1pt}{0.6mm} } & 1,000  & 15 & $2\pi $ & 9.5 & 4 & 0.06 & 5.63 &1,792 &270 & 1,792 & 25.9 &  2nd order FD  \cite{ste13}\\
	  CHJ020 \cite{hoy06} & Channel & \crline{green} & 2,000  & $8\pi$ & $3\pi$ & $8.1$ & $4.1 $ & 0.004 & 8.9 & 6,144 & 633 & 4,608 & 10.3 &  PS + CFD \cite{llu21c}\\
         PYR020 \cite{yao23} & Pipe & \prline{yellow} & 2,000  & $10\pi$ & $2\pi $ & $10.2$ & $3.9$ & 0.1 & 4.9 & 6,144 & 768 & 2,560  & 9.7 & PS + HOFD \cite{wil17}\\
        PPR020 \cite{pir21} & Pipe & {\color{yellow}\hdashrule[0.75ex]{0.86cm}{1pt}{0.6mm} } & 2,000  & 15 & $2\pi $ & 9.7 & 4 & 0.06 & 6.61 &3,072 &399 & 3,072 & 22.4 & 2nd order FD \cite{ste13}\\
	CLM050 \cite{lee15} & Channel  &  \crline{gray} & 5,200 & $8\pi $ & $3\pi $ & $12.7$ & $6.4$ & 0.5 & $ 10.3$ &10,240 & 1,536 & 7,680 & 7.80 & PS + Splines \cite{lee14} \\ 
         PYR050 \cite{yao23} & Pipe & \prline{blue} & 5,000  & $10\pi$ & $2\pi $ & $12.8$ & $6.3$ & 0.2 & 8.6 & 12,288 & 1,024 & 5,120  & 4.6 & PS + HOFD \cite{wil17}\\
        PPR060 \cite{pir21} & Pipe & {\color{blue}\hdashrule[0.75ex]{0.86cm}{1pt}{0.6mm} } & 6,000  & 15 & $2\pi $ & 9.7 & 4 & 0.06 & 6.61 &9,216 &910 & 9,216 & 8.3 &  2nd order FD  \cite{ste13}\\
  	CKY080 \cite{kan21} & Channel  &  \crline{black}  & 8,000 & $16$ & $6.4$ & $12.3 $ & $5.9 $ & 0.6 & 8.0 & 10,368 & 4,096 & 8,640 & 6.3 & PS + 2nd order FD \cite{yam18}\\ 
	CHO100 \cite{hoy22} & Channel  &  \crline{yellow}& 10,000 & $2\pi $ & $\pi $ & $10.1$ & $5.2$ & 0.3 & 13 & 6,144 & 2,101 & 6,144 & 19.8 & PS + CFD \cite{llu21c} \\
   	CYT160 \cite{yam24} & Channel &   \crline{blue} & 16,000 &  16 & $ 6.4 $ & $11.85$ & 4.7& $0.6$ & 12.4 & 21,600 & 5,760 & 20,736 & 6.0 & PS + 2nd order FD \cite{yam18}\\

	\end{tabular}
		\caption{Parameters of DNS cases used in various figures, and listed here in order of increasing friction Reynolds number, $\Delta_x^+$ and $\Delta_z^+$ are in terms of dealiased Fourier modes (physical space), and $y^+$ is the inner-scaled wall distance.  The next to last column is the total simulation time without initial transients in terms of eddy turnovers. The last column indicates the method used by the authors, where, PS: pseudo spectral \cite{can12}, FD: finite differences, and CFD: compact finite differences \cite{lel92}}
		\label{tab:1}
	\end{center}
\end{table*}